\documentclass[prb,showpaces,preprintnumbers,amsmath,amssymb,twocolumn,superscriptaddress]{revtex4-1}
\usepackage{graphicx}
\usepackage{dcolumn}
\usepackage{bm}
\usepackage{amssymb}
\usepackage{amsmath}
\usepackage{epstopdf}
\begin{document}

\title{Tunable interplay between $3d$ and $4f$ electrons in Co-doped iron pnictides}

\author{T. Shang}
\affiliation{Department of Physics and Center for Correlated Matter, Zhejiang University, Hangzhou, 310027, China}
\author{L. Yang}
\affiliation{Department of Physics and Center for Correlated Matter, Zhejiang University, Hangzhou, 310027, China}
\author{Y. Chen}
\affiliation{Department of Physics and Center for Correlated Matter, Zhejiang University, Hangzhou, 310027, China}
\author{N. Cornell}
\affiliation{UTD-NanoTech Institute, The University of Texas at Dallas, Richardson, Texas 75083$\textendash$0688, USA}
\author{F. Ronning}
\affiliation{Los Alamos National Laboratory, Los Alamos, New Mexico 87545, USA}
\author{J. L. Zhang}
\affiliation{Department of Physics and Center for Correlated Matter, Zhejiang University, Hangzhou, 310027, China}
\author{L. Jiao}
\affiliation{Department of Physics and Center for Correlated Matter, Zhejiang University, Hangzhou, 310027, China}
\author{Y. H. Chen}
\affiliation{Department of Physics and Center for Correlated Matter, Zhejiang University, Hangzhou, 310027, China}
\author{J. Chen}
\affiliation{Department of Physics and Center for Correlated Matter, Zhejiang University, Hangzhou, 310027, China}
\author{A. Howard}
\affiliation{UTD-NanoTech Institute, The University of Texas at Dallas, Richardson, Texas 75083$\textendash$0688, USA}
\author{J. Dai}
\affiliation{Condensed Matter Group, Department of physics, Hangzhou Normal University, Hangzhou 310036, China}
\author{J. D. Thompson}
\affiliation{Los Alamos National Laboratory, Los Alamos, New Mexico 87545, USA}
\author{A. Zakhidov}
\affiliation{UTD-NanoTech Institute, The University of Texas at Dallas, Richardson, Texas 75083$\textendash$0688, USA}
\author{M. B. Salamon}
\affiliation{UTD-NanoTech Institute, The University of Texas at Dallas, Richardson, Texas 75083$\textendash$0688, USA}
\author{H. Q. Yuan}
\email{hqyuan@zju.edu.cn}
\affiliation{Department of Physics and Center for Correlated Matter, Zhejiang University, Hangzhou, 310027, China}
\date{\today}
\begin{abstract}
We study the interplay of $3d$ and $4f$ electrons in the iron pnictides CeFe$_{1-x}$Co$_x$AsO and GdFe$_{1-y}$Co$_y$AsO, which correspond to two very different cases of $4f$-magnetic moment. Both CeFeAsO and GdFeAsO undergo a spin-density-wave (SDW) transition associated with Fe $3d$ electrons at high temperatures, which is rapidly suppressed by Fe/Co substitution. Superconductivity appears in a narrow doping range: $0.05 < x < 0.2$ for CeFe$_{1-x}$Co$_x$AsO and $0.05 < y < 0.25$ for GdFe$_{1-y}$Co$_y$AsO, showing a maximum transition temperature $T_\textup{sc}$ of about 13.5 K for Ce and 19 K for Gd. In both compounds, the $4f$-electrons form an antiferromagnetic (AFM) order at low temperatures over the entire doping range and Co $3d$ electrons are ferromagnetically ordered on the Co-rich side; the Curie temperature reaches $T_\textup{C}^\textup{Co} \approx$ 75 K at $x = 1$ and $y = 1$. In the Ce-compounds, the N\'{e}el temperature $T_\textup{N}^\textup{Ce}$ increases upon suppressing the SDW transition of Fe and then remains nearly unchanged with further increasing Co concentration up to $x \simeq 0.8$ ($T_\textup{N}^\textup{Ce}\approx$ 4 K). Furthermore, evidence of Co-induced polarization on Ce-moments is observed on the Co-rich side. In the Gd-compounds, the two magnetic species of Gd and Co are coupled antiferromagnetically to give rise to ferrimagnetic behavior in the magnetic susceptibility on the Co-rich side. For $0.7 \leq y < 1.0$, the system undergoes a possible magnetic reorientation below the N\'{e}el temperature of Gd ($T_\textup{N}^\textup{Gd}$). Our results suggest that the effects of both electron hybridizations and magnetic exchange coupling between the $3d$-$4f$ electrons give rise to a rich phase diagram in the rare-earth iron pnictides.
\begin{description}
\item[PACS number(s)]
74.70.Xa,71.20.Eh,75.20.Hr
\end{description}
\end{abstract}
\maketitle

\section{\label{sec:level1}INTRODUCTION}

The discovery of superconductivity at $T_\textup{sc}$ = 26 K in LaFeAsO$_{1-x}$F$_x$ has stimulated intensive efforts on searching for new materials with higher superconducting transition temperature, $T_\textup{sc}$, and revealing their unconventional nature.~\cite{kamihara2008iron} Until now, a few series of iron-based superconductors (FeSCs) have been discovered, with a maximum $T_\textup{sc}$ raised to 56 K when La is substituted by other rare-earth elements, e.g., Ce, Sm, Nd, Pr and Gd, in the $Re$Fe$Pn$O family ($Re$ = rare earth, $Pn$= pnictogen).~\cite{cxh2008, chen2008superconductivity, Ren2008, wang2008thorium} The FeSCs show strong similarities to the copper oxides:~\cite{armitage2010progress, reviews} (i) both possess a layered crystal structure and a relative high $T_\textup{sc}$; (ii) superconductivity seems to be closely tied up with magnetism. On the other hand, significant differences have been observed between them. For example, the parent compounds of FeSCs are bad metals, in contrast to the Mott-insulators in the cuprates.~\cite{armitage2010progress, reviews} The superconducting order parameter of FeSCs is supposed to be s$_\pm$-type,~\cite{mazin2008unconventional,Mazinreview} while the cuprates are d-wave superconductors.~\cite{Harlingen, Tsuei} Moreover, the FeSCs show nearly isotropic upper critical fields at low temperature,~\cite{yuan2009nearly,ZhangHC2} but the cuprates are rather anisotropic.~\cite{worthington1987anisotropic} In these compounds, elemental substitutions have been shown to be an efficient approach to study superconductivity and its interplay with magnetism.

In $Re$FeAsO, superconductivity can be induced by doping either electrons or holes into the $Re$O layers, as well as doping electrons into the FeAs layers. Upon substituting Fe with Co, the magnetic/structural transition is quickly suppressed and superconductivity exists over a narrow doping range.~\cite{zhao2010effects,sefat2008superconductivity,shirage2009search,marcinkova2010superconductivity,wang2009effects} In the $Re$O layers, partial substitution of Gd with Th suppresses the magnetic/structural transition and then gives rise to superconductivity where $T_\textup{sc}$ reaches a maximum of 56 K at 20\% Th.~\cite{wang2008thorium} Similarly, a narrow superconducting region was also observed in GdFeAsO$_{1-\delta }$ by introducing oxygen deficiency.~\cite{yang2008superconductivity} Furthermore, O/F substitution in CeFeAsO also induces superconductivity with $T_\textup{sc}$ raised to 41 K.~\cite{chen2008superconductivity}

On the other hand, the rare-earth elements usually form an antiferromagnetically ordered state at very low temperatures. For instance, CeFeAsO sequentially undergoes two AFM-type transitions upon cooling down from room temperature, one associated with Fe ($T_\textup{N}^\textup{Fe} \approx $ 150 K) and the other one attributed to Ce ($T_\textup{N}^\textup{Ce} \approx $ 3.4 K).~\cite{chen2008superconductivity, zhao2008structural} A similar situation happens in GdFeAsO ($T_\textup{N}^\textup{Fe} \approx $ 130 K, $T_\textup{N}^\textup{Gd} \approx $ 4.2 K).~\cite{wang2008thorium} On the other hand, CeFePO is a paramagnetic(PM) heavy fermion metal.~\cite{E.M.Bruning2008cefepo} Recent studies demonstrated that a ferromagnetic (FM) state of Ce $4f$ electrons, separating the AFM state from the PM heavy fermion state, develops in the intermediate doping region in CeFeAs$_{1-x}$P$_{x}$O.~\cite{luo2010phase, de2010lattice, Jesche} Moreover, the Co-end compounds, namely $Re$Co$Pn$O, demonstrate rich magnetic properties attributed to the interactions between the $3d$ and $4f$ electrons. For example, CeCoAsO and CeCoPO exhibit an enhanced Sommerfeld coefficient of 200 mJ/mol-K$^{2}$, and their Co-ions undergo a FM transition at $T_\textup{C}^\textup{Co}$ = 75 K.~\cite{sarkar2010interplay, krellner2009interplay} LaCoAsO and LaCoPO are itinerant ferromagnets with a Curie temperature of $T_\textup{C}^\textup{Co}=$ 50 K and 60 K, respectively.~\cite{yanagi2008itinerant} Complex magnetic properties were observed in SmCoAsO and NdCoAsO, in which the Co-ions first undergo a FM transition and then an AFM transition upon cooling from room temperature,~\cite{awana2010magnetic, mcguire2010magnetic} followed by an AFM transition of the $4f$ electrons at $T_{\textup{N}}^{\textup{Sm}}$ = 5 K and $T_{\textup{N}}^{\textup{Nd}}$ = 3.5 K.~\cite{awana2010magnetic, mcguire2010magnetic} In GdCoAsO, both Co and Gd are magnetically ordered too,~\cite{ohta2009magnetic} but there is no systematic study so far. These indicate that the 1111-type iron pnictides may provide us a rare system to study the interplay of $3d$ and $4f$ electrons, and their emergent properties. A systematic study of the interplay between $3d$ and $4f$ electrons would help elucidate the nature of superconductivity and magnetism in iron pnictides.

In this paper, we synthesize and measure the physical properties of CeFe$_{1-x}$Co$_x$AsO and GdFe$_{1-y}$Co$_y$AsO ($0 \leq x,y \leq 1$), which represent two different cases of $4f$ magnetic moments. A complete temperature-doping phase diagram is derived for each compound, showing remarkably rich physical properties tuned by $3d$-$4f$ interactions. The article is organized as follows: Immediately after the introduction, we describe our experimental methods in section II. Section IIIA characterizes the crystal structure of the compounds. In section IIIB, C and D, we present the experimental results of the electrical resistivity, magnetic properties and specific heat for CeFe$_{1-x}$Co$_x$AsO and GdFe$_{1-y}$Co$_y$AsO, from which their phase diagrams are derived (section IIIE). Finally, a summary is given in section IV.

\section{EXPERIMENTAL DETAILS}

\begin{figure}[tbp]
\begin{center}
\includegraphics[width=3.2in,keepaspectratio]{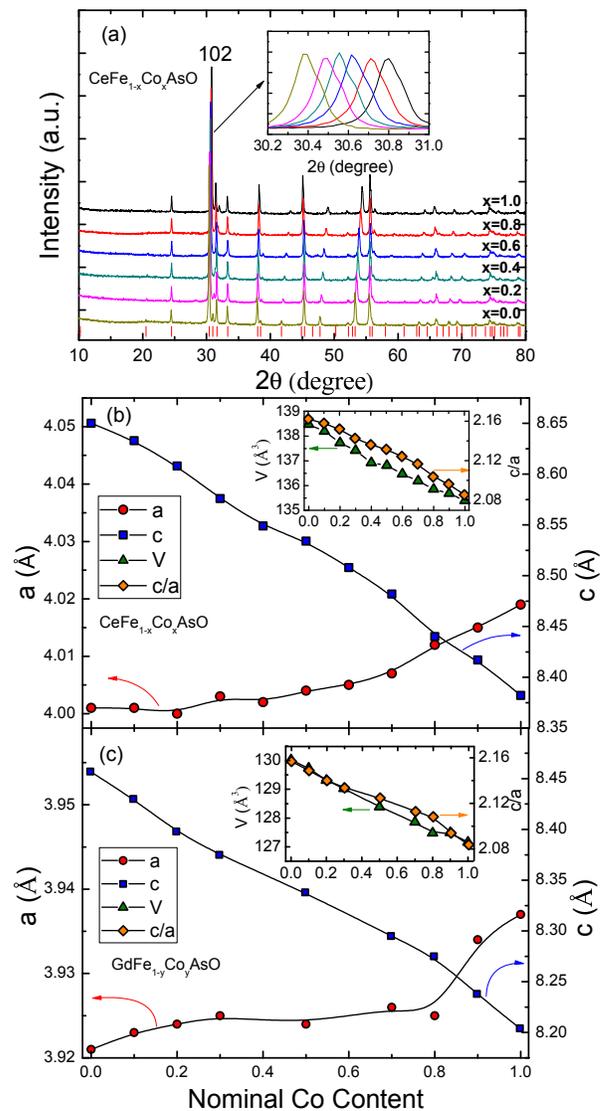}
\end{center}
\caption{(Color online) (a) Room temperature powder X-ray diffraction patterns of CeFe$_{1-x}$Co$_x$AsO with $x = 0.0, 0.2, 0.4, 0.6, 0.8, 1.0$. The inset shows the enlarged (\textbf{102}) peaks. (b) and (c) Lattice parameters as a function of nominal Co content for CeFe$_{1-x}$Co$_x$AsO and GdFe$_{1-y}$Co$_y$AsO, respectively. The insets plot the unit cell volume (left axis) and the $c/a$ ratio (right axis) as a function of doping.}
\label{fig1}
\end{figure}

Polycrystalline samples of CeFe$_{1-x}$Co$_{x}$AsO and GdFe$_{1-y}$Co$_{y}$AsO ($0\leq x,y\leq 1$) were synthesized in evacuated quartz ampoules by a two-step solid state reaction method. High-purity Ce (99.8$\%$), Gd (99.9$\%$) and As (99.999$\%$) chunks, also Co$_{3}$O$_{4}$ (99.97$\%$), Fe$_{2}$O$_{3}$ (99.99$\%$), Fe (99.9$\%$) and Co (99.9$\%$) powders were used as raw materials. In the following, we describe the methods of synthesizing the CeFe$_{1-x}$Co$_{x}$AsO compounds; the same holds for the Gd-compounds. Dehydrated Co$_{3}$O$_{4}$ and Fe$_{2}$O$_{3}$ were prepared by heating the powders at 1173 K for 10 h. First, CeAs precursor was prepared by heating the mixture in a sealed evacuated quartz ampoule at 1073 K for 24 h, and then at 1323 K for 48 h. Then, CeAs, Co$_{3}$O$_{4}$, Fe$_{2}$O$_{3}$, Fe and Co powders were weighed according to the stoichiometric ratio, thoroughly ground, and pressed into pellets. The entire process was carried out in an argon-filled glove box. Finally, the pellets were placed in alumina crucibles, sealed into evacuated quartz ampoules, and then slowly heated to 1323 K at a rate of 100 K/h and annealed at this temperature for one week to obtain sintered pellets.

The structure of these polycrystalline samples was characterized by powder X-ray diffraction (XRD) at room temperature using a PANalytical X'Pert MRD diffractometer with Cu K$\alpha $ radiation and a graphite monochromator. Measurements of the DC magnetic susceptibility, electrical resistivity and specific heat were carried out in a Quantum Design Magnetic Property Measurement System (MPMS-5) and a Physical Property Measurement System (PPMS-9), respectively. Temperature dependence of the electrical resistivity was measured by a standard four-point method.

\section{RESULTS AND DISCUSSION}
\subsection{Crystal Structure}
Figure 1(a) plots several representative XRD patterns of the CeFe$_{1-x}$Co$_{x}$AsO polycrystalline samples. The GdFe$_{1-y}$Co$_{y}$AsO compounds show similar XRD patterns (not show here). The vertical bars on the bottom denote the theoretically calculated positions of the Bragg diffractions for CeFeAsO. All the peaks can be well indexed based on the tetragonal ZrCuSiAs structure with a space group of P4/nmm. No obvious impurity phases are detected, indicating high quality of these samples. The shift of the (\textbf{102}) peak toward the right (larger $2\theta $ value) with increasing Co concentration, as shown in the right inset of Fig.1(a), reveals a contraction of the crystal lattice with increasing Co-concentration. The lattice parameters, refined by the least square fitting, are plotted as a function of nominal Co-concentration in Fig.1(b) and Fig.1(c) for CeFe$_{1-x}$Co$_{x}$AsO and GdFe$_{1-y}$Co$_{y}$AsO, respectively. Different from CeFeAs$_{1-x}$P$_{x}$O where both the $a$- and $c$-axis shrink with increasing P-content,~\cite{luo2010phase} in CeFe$_{1-x}$Co$_{x}$AsO and GdFe$_{1-y}$Co$_{y}$AsO, the $c$-axis shrinks significantly while the $a$-axis increases slightly with increasing Co concentration, meaning that Fe/Co substitution increases three dimensionality in comparison with As/P substitution. This becomes more pronounced for $x \geq 0.75$ (Ce) and  $y \geq 0.8$ (Gd), in which regime Co-electrons order ferromagnetically at low temperatures. The shrinkage of the $c$-axis reduces the distance between the CeO(GdO) and FeAs/CoAs layers, and thus enhances the hybridizations and magnetic exchange interactions between $3d$ and $4f$ electrons. The inset of Fig.1(b) and Fig.1(c) shows the unit cell volume and the $c/a$ ratio versus Co concentration, both of which monotonically decrease with increasing Co content. The former decreases by 2.22$\%$ and 2.2$\%$ at $x = 1$ and $y = 1$, respectively. The shift of the XRD patterns and the variation of lattice parameters suggest that the Co atoms are successfully incorporated into the crystal lattice.

\subsection{Electrical Resistivity}

\subsubsection{CeFe$_{1-x}$Co$_x$AsO}

\begin{figure}[tbp]
\begin{center}
\includegraphics[width=3.2in,keepaspectratio]{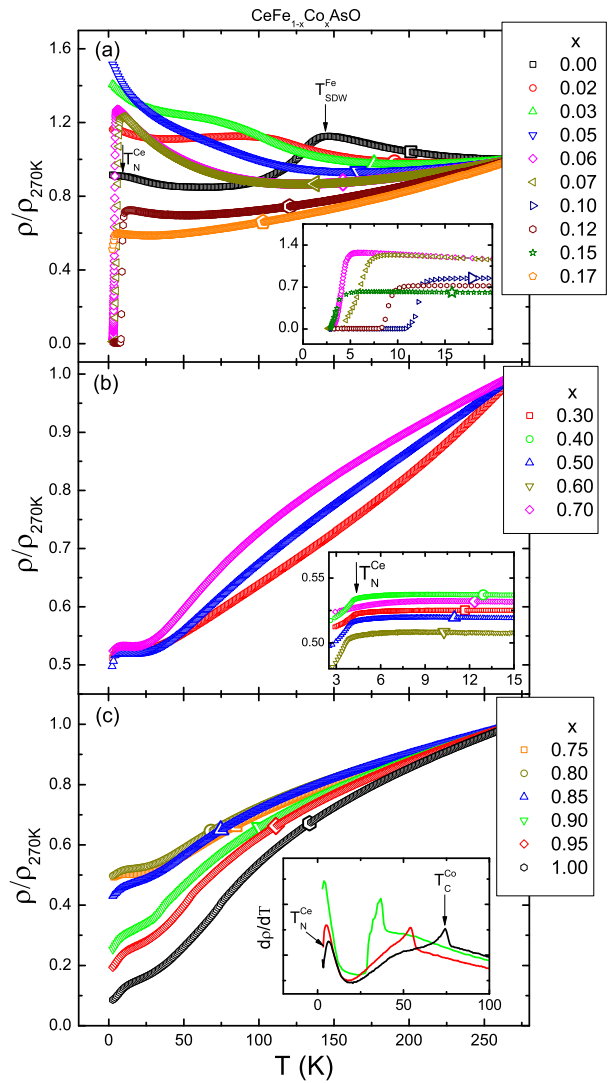}
\end{center}
\caption{(Color online) Temperature dependence of the normalized electrical resistivity of CeFe$_{1-x}$Co$_{x}$AsO. (a) $0 \leq x \leq 0.2$. (b) $0.3 \leq x \leq 0.7$. (c) $0.75 \leq x \leq 1$. The insets of (a) and (b) expand the low temperature regime. The inset of (c) shows the derivative plots of the electrical resistivity with respect to temperature for $x$ = 0.9, 0.95 and 1, from which the transition temperatures of $T_\textup{N}^\textup{Ce}$ and $T_\textup{C}^\textup{Co}$ can be determined.}
\label{fig2}
\end{figure}

The temperature dependence of the normalized electrical resistivity is shown in Fig.2 for CeFe$_{1-x}$Co$_{x}$AsO ($0 \leq x \leq 1$). The resistive anomaly observed around $T_\textup{SDW}^\textup{Fe} \approx$ 150 K in CeFeAsO characterizes the almost coincident SDW-transition of Fe $3d$ electrons and the tetragonal to orthorhombic structural phase transition.~\cite{chen2008superconductivity,zhao2008structural,dong2008competing} With further decreasing temperature, another resistive kink appears around $T_\textup{N}^\textup{Ce}\approx 4$ K, marking the AFM transition of Ce.~\cite{chen2008superconductivity, zhao2008structural} Upon partially substituting Fe with Co, the transition at $T_\textup{SDW}^\textup{Fe}$ is quickly suppressed and becomes hardly visible above $x \approx 0.06$. Meanwhile, superconductivity appears around $x = 0.06$, and the transition temperature reaches a maximum of $T_\textup{sc}^\textup{onset} \approx $ 13.5 K at $x = 0.1$ [see the inset of Fig.2(a)]. Note that, for $0.06\leq x\leq 0.17$, no signature of a magnetic transition can be identified below $T_\textup{sc}$ in the electrical resistivity. This is different from that of CeFeAsO$_{1-x}$F$_{x}$ in which a non-zero resistive transition shows up below $T_\textup{sc}$ attributed to the magnetic order of Ce.~\cite{tianCe} However, evidence for the AFM transition of Ce can be inferred from the magnetic susceptibility and specific heat (see below). In comparison with CeFeAsO$_{1-x}$F$_{x}$,~\cite{chen2008superconductivity} another difference is the low-temperature semiconducting behavior observed in the underdoped CeFe$_{1-x}$Co$_{x}$AsO ($x < 0.1$), which may originate from a disorder effect or Kondo-like scattering. The tiny amount of Co atoms in the materials may act as magnetic impurities, leading to a resistive minimum as a result of the Kondo effect. The system eventually becomes metallic with further increasing Co concentration. Figure 2(b) presents the normalized electrical resistivity for $0.3\leq x\leq 0.7$, which decreases monotonically with decreasing temperature. In this doping region, only one pronounced resistive transition is seen around 4 K, which is attributed to the AFM transition of Ce $4f$ electrons, as shown by the arrow in the inset of Fig.2(b). For $x \geq 0.75$, the electrical resistivity demonstrates stronger temperature dependence with a broad hump around 100 K, likely attributed to the enhanced $3d$-$4f$ hybridizations as typically found in some Kondo lattice compounds.~\cite{CeRh2Si2, Ce2Rh3Ge5} Upon further Co-doping, the low-temperature plateau, observed for $0.3\leq x\leq 0.7$, persists as a change in slope as clearly shown in the derivatives of the electrical resistivity with respect to temperature, $d\rho /dT$, in the inset of Fig.2(c). One peak corresponds to the FM transition of Co ($T_\textup{C}^\textup{Co}$) and the other, to the AFM order of Ce ($T_\textup{N}^\textup{Ce}$), both shifting to higher temperature with further increasing Co concentration. The magnetic transition of Ce is broadened on the Co-rich side; we attribute it to the polarization effect of Co-ferromagnetism which will be further illustrated in terms of magnetic susceptibility and specific heat.

\subsubsection{GdFe$_{1-y}$Co$_y$AsO}
\begin{figure}[tbp]
\begin{center}
\includegraphics[width=3.35in,keepaspectratio]{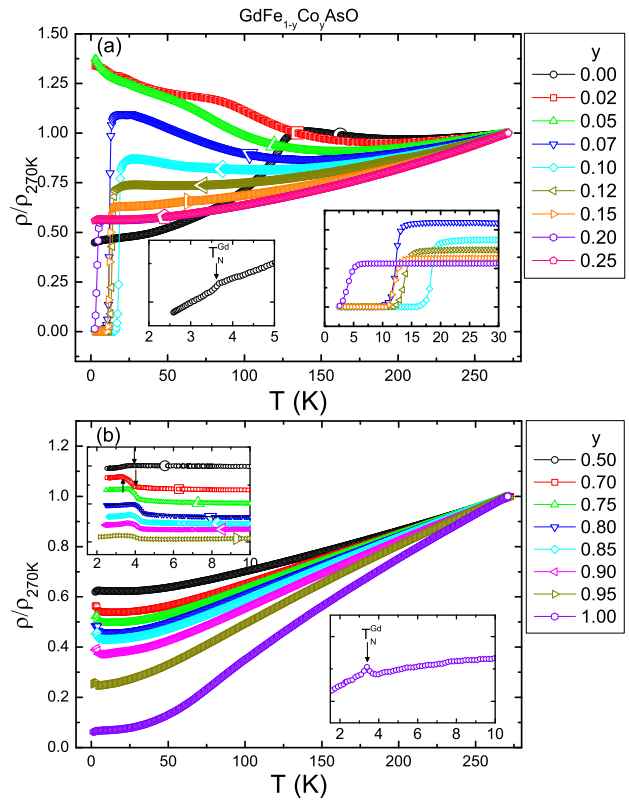}
\end{center}
\caption{(Color online) Temperature dependence of the normalized electrical resistivity of GdFe$_{1-y}$Co$_y$AsO. (a) $0.0 \leq y \leq 0.25$. (b) $0.5
\leq y \leq 1.0$. The insets expand the low temperature regime.}
\label{fig3}
\end{figure}

\begin{figure}[tbp]
\begin{center}
\includegraphics[width=3.35in,keepaspectratio]{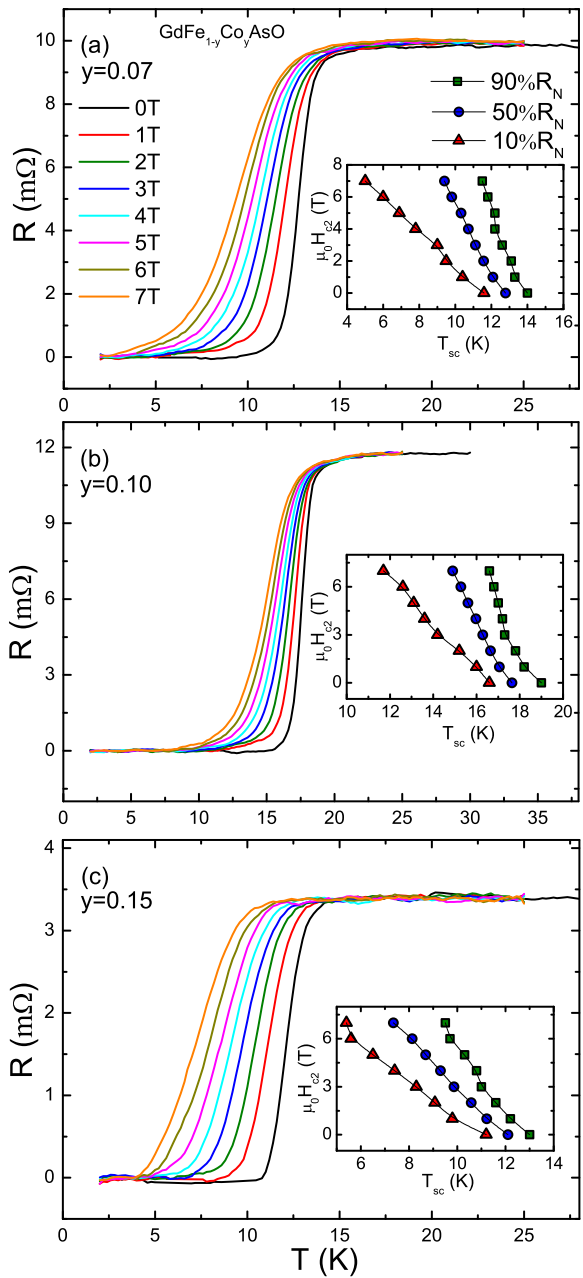}
\end{center}
\caption{(Color online) Temperature dependence of the electrical resistance at various magnetic fields for $y$ = 0.07 (a), 0.10 (b), 0.15 (c). The insets show the upper critical fields $\mu_0 H_\textup{c2}(T_\textup{sc})$ obtained at the 90$\%$ (green square), 50$\%$ (blue circle), 10$\%$ (red triangle) of the normal resistance R$_\textup{N}$.}
\label{fig4}
\end{figure}

Figure 3 plots the temperature dependence of the normalized electrical resistivity for GdFe$_{1-y}$Co$_{y}$AsO ($0 \leq y \leq 1$) at zero-field. In GdFeAsO, the coincident Fe-SDW transition and the structural phase transition decreases to $T_\textup{SDW}^\textup{Fe} \approx $ 130 K.~\cite{wang2008thorium, zhao2008structural, dong2008competing} At very low temperatures, a resistive kink appears around 3.6 K, as shown by the arrow in the left inset of Fig.3(a), below which the Gd $4f$ electrons become antiferromagnetically ordered.~\cite{wang2008thorium} The $T_\textup{SDW}^\textup{Fe}$ is quickly suppressed upon substituting Fe with Co and becomes imperceptible around $y = 0.07$. Simultaneously, superconductivity shows up around $y$ = 0.07, with the transition temperature reaching a maximum of $T_\textup{sc}^\textup{onset} \approx 19$ K around $y = 0.1$ [see the right inset of Fig.3(a)]. Similar to the Ce-compounds, the Gd-AFM transition is not visible in the electrical resistivity of the superconducting samples ($0.07 \leq y \leq 0.2$), but it can be inferred from the magnetic susceptibility (see below). Upon further increasing the Co concentration, no superconductivity can be observed down to 3 K, and the AFM transition of Gd shows up around 4 K for $y = 0.25$. In the underdoped GdFe$_{1-y}$Co$_{y}$AsO ($y < 0.1$), the electrical resistivity also shows semiconducting behavior.

Figure 3(b) presents the normalized electrical resistivity for the heavily Co-doped samples ($0.5\leq y\leq 1.0$), which decreases monotonically with decreasing temperature, showing metallic behavior above 5 K. The electrical resistivity demonstrates a stronger temperature dependence for $y \geq 0.85$, which may be attributed to spin-wave scattering from the FM Co. Again, the Gd $4f$ electrons undergo an AFM transition below 4 K. In particular, a step-like upturn transition appears at low temperature for $0.7 \leq y \leq 0.95$, which can be clearly seen from the upper inset of Fig.3(b). For example, the electrical resistivity for $y = 0.7$ increases abruptly at 3.9 K and then decreases below 3.2 K. Such a resistive upturn might be associated with a gap opening at the AFM transition temperature of Gd. The subsequent resistive decrease may arise from a magnetic reorientation, which is observed in the bulk properties too (see below). This step-like transition disappears in the stoichiometric GdCoAsO material, in which only a small kink associated with the AFM transition of Gd $4f$ electrons is observed around $T_\textup{N}^\textup{Gd}$ $\approx $ 3.5 K, as shown in the lower inset of Fig.3(b).

Figure 4 presents the resistive superconducting transitions for $y$ = 0.07, 0.10, and 0.15 at various magnetic fields. In zero field, all the samples show a sharp transition with $T{^\textup{{mid}}_\textup{sc}}$ $\approx$ 12.6 K, 17.6 K and 12.2 K for $y$ = 0.07, 0.10 and 0.15, respectively. Application of a magnetic field shifts the superconducting transition to lower temperatures and significantly broadens the transition width. The latter is likely attributed to the vortex flow as previously discussed in other Fe-based superconductors.~\cite{jiao, aroszynski, Lee} The insets of Fig.4 plot the upper critical fields $\mu_0 H_\textup{c2}(T_\textup{sc})$, which were derived at various resistive drops of the superconducting transition. Following the WHH model: $\mu_0 H_{c2}(0)=-0.693T_\textup{sc}[d\mu_0H_\textup{c2}/dT]$,~\cite{werthamer1966temperature} one can estimate $\mu_0$$H_\textup{c2}$(0), which gives $\mu_0$$H_\textup{c2}$(0) = 19 T, 32 T and 13 T for $y$ = 0.07, 0.1 and 0.15, respectively. The coherence length $\xi(0)$ can be calculated from the Ginzburg-Landau formula $\xi(0) \approx (\Phi_{0}/2 \pi \mu_0 H_{c2})^{1/2}$, where $\Phi_0$ is the quantum of magnetic flux. This yields $\xi(0) = 41 \textup{\r{A}}, 32 \textup{\r{A}}$ and 50 $\textup{\r{A}}$ for $y = 0.07, 0.10$ and 0.15, respectively.

\subsection{Magnetic Properties}

\subsubsection{CeFe$_{1-x}$Co$_x$AsO}

\begin{figure}[tbp]
\begin{center}
\includegraphics[width=3.2in,keepaspectratio]{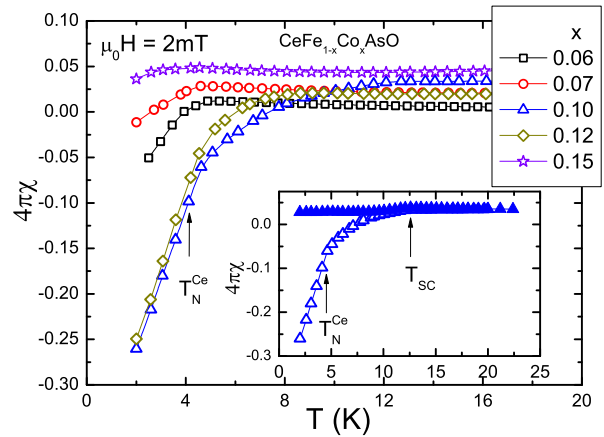}
\end{center}
\caption{(Color online) Low temperature magnetic susceptibility $\protect\chi(T)$ for CeFe$_{1-x}$Co$_x$AsO, $0.06 \leq x \leq 0.15$. The open and filled symbols show the data measured at zero-field cooling (ZFC) and field cooling (FC), respectively. To demonstrate the existence of an AFM transition of Ce below $T_\textup{sc}$, we plot $\protect\chi(T)$ for $x = 0.1$ in the inset, from which a kink at $T_\textup{N}^\textup{Ce}$ can be identified.}
\label{fig5}
\end{figure}

\begin{table}[tbp]
\caption{The superconducting transition temperatures $T_\textup{sc}$ (K) for CeFe$_{1-x}$Co$_x$AsO, $0.06 \leq x \leq 0.17$, determined from 90$\%$ ($T{^\protect\rho_\textup{sc}}$ (onset)), 50$\%$ ($T{^\protect\rho_\textup{sc}}$ (mid)) and 10$\%$ ($T{^\protect\rho_\textup{sc}}$ (zero)) of the normal resistivity at $T_\textup{sc}$ and the onset of magnetic susceptibility $T{^\protect\chi_\textup{sc}}$.\newline}
\label{tab:table1}\centering
\begin{ruledtabular}
\begin{tabular}{lcccc}
\textrm{$x$}&
\textrm{$T{^\rho_\textup{sc}}$ (onset)}&
\textrm{$T{^\rho_\textup{sc}}$ (mid)}&
\textrm{$T{^\rho_\textup{sc}}$ (zero)}&
\textrm{$T{^\chi_\textup{sc}}$}\\
\colrule
0.06 & 4.7 & 3.95 & 3.2 & 4.6\\
0.07 & 7.3 & 5.9 & 4.4 & 4.6\\
0.10 & 13.5 & 12 & 11.1 & 13\\
0.12 & 10 & 9.2 & 8.4 & 9.1\\
0.15 & 4.4 & 3.6 & 2.9 & 2.85\\
0.17 & 4.3 & \\
\end{tabular}
\end{ruledtabular}
\end{table}

The dc magnetization of CeFe$_{1-x}$Co$_x$AsO was measured as functions of temperature and magnetic field. Figure 5 shows the magnetic susceptibility $\chi(T)$ for the superconducting samples, measured in a field of $\mu_0 H = 2 $ mT. Evidence of bulk superconductivity was observed for $0.06 \leq x \leq 0.15$. The superconducting volume fraction reaches over $25\%$ for $x = 0.1$ and $x = 0.12$. Furthermore, a kink can be tracked around $T_\textup{N}^\textup{Ce} \approx 4$ K at various doping concentrations (see the arrows for $x$ = 0.1 and 0.12). Likely, this corresponds to the AFM transition of Ce in the superconducting state. It is noted that such a magnetic transition of Ce is clearly demonstrated in the specific heat (see below). The superconducting transition temperatures $T_\textup{sc}$, derived from the electrical resistivity and magnetic susceptibility, are summarized in Table I.

\begin{figure}[tbp]
\begin{center}
\includegraphics[width=3.2in,keepaspectratio]{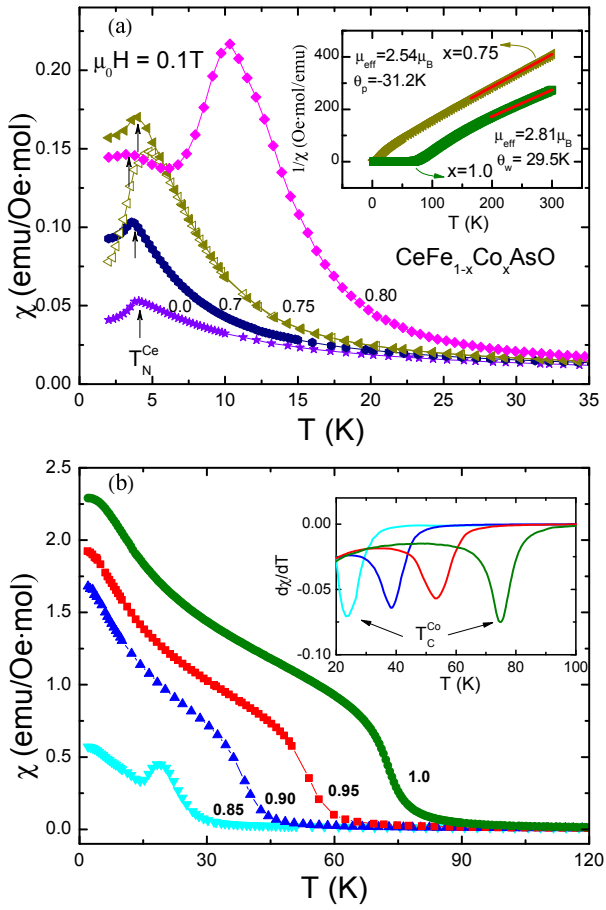}
\end{center}
\caption{(Color online) Temperature dependence of the DC magnetic susceptibility for CeFe$_{1-x}$Co$_x$AsO, $0 \leq x \leq 1$. The open and filled symbols present the ZFC and FC data, respectively. The upper inset plots the inverse susceptibility as a function of temperature for $x = 0.75$ and $x = 1$. The red lines are fits to a Curie-Weiss law. The Curie temperature $T_\textup{C}^\textup{Co}$ of Co is determined from the negative peak position of the derivative d$\protect\chi(T)$/d$T$ as shown in the lower inset.}
\label{fig6}
\end{figure}

\begin{figure}[tbp]
\begin{center}
\includegraphics[width=3.2in,keepaspectratio]{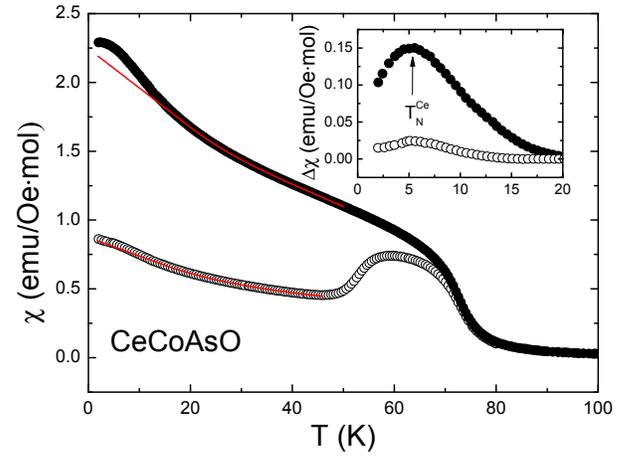}
\end{center}
\caption{(Color online) Temperature dependence of the magnetic susceptibility $\protect\chi(T)$ for CeCoAsO. The open and filled circles show the ZFC and FC data, respectively. The red lines present a polynomial fit to the experimental data in the temperature range of 20 K to 40 K, which are treated as the FM background of Co. The inset plots the magnetic susceptibility after subtracting the FM background.}
\label{fig7}
\end{figure}

\begin{figure*}[tbp]
\begin{center}
\includegraphics[width=7in,keepaspectratio]{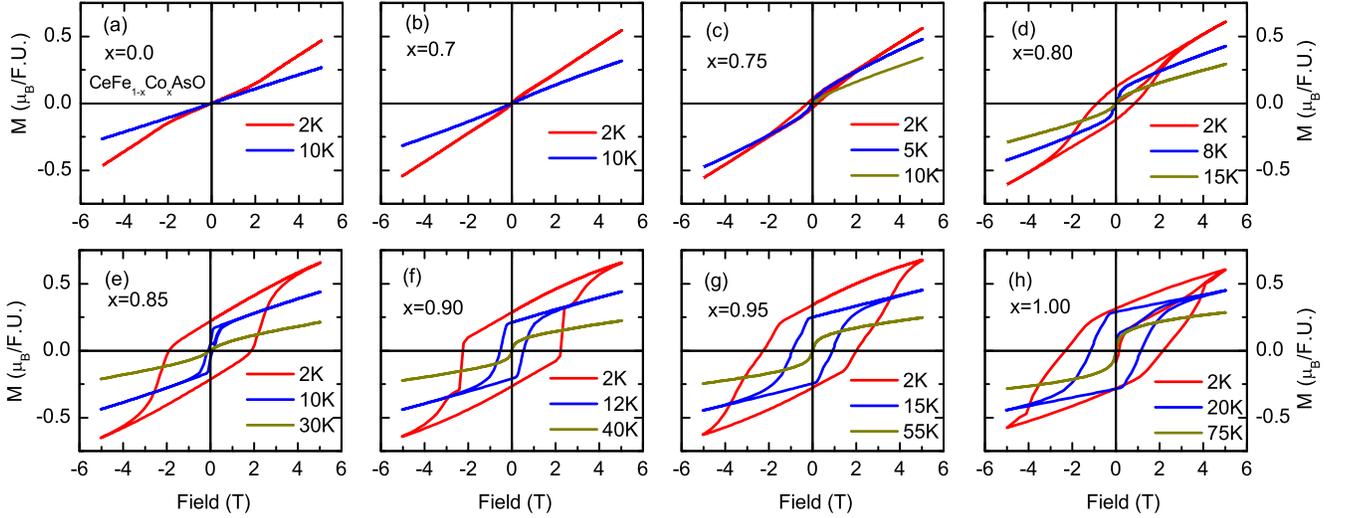}
\end{center}
\caption{(Color online) Field dependence of the magnetization $M(H$) at various temperatures for CeFe$_{1-x}$Co$_x$AsO, $0 \leq x \leq 1$. The dark yellow curves are measured around the Curie temperature $T^\textup{Co}_\textup{C}$.}
\label{fig8}
\end{figure*}

\begin{figure}[tbp]
\begin{center}
\includegraphics[width=3.2in,keepaspectratio]{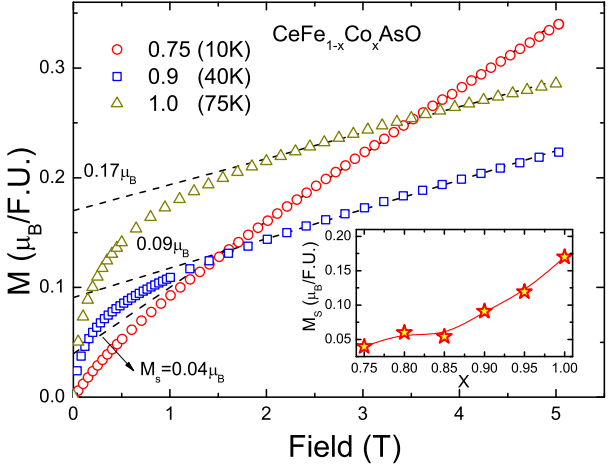}
\end{center}
\caption{(Color online) $M(H)$ curves around the Curie temperature $T^\textup{Co}_\textup{C}$ for CeFe$_{1-x}$Co$_x$AsO, $x = 0.75, 0.9$ and $1$. The dashed lines describe the methods of deriving the saturated moments. The inset plots the saturated moments as a function of Co concentration.}
\label{fig9}
\end{figure}

Magnetic susceptibility of the non-superconducting samples, measured in a magnetic field of $\mu_0 H$ = 0.1 T, is presented in Fig.6. Over a wide doping range of $0 \leq x \leq 0.8$, the magnetic susceptibility $\chi(T)$ consistently shows a pronounced peak around $T^\textup{Ce}_\textup{N}\approx$ 4 K, which is attributed to the AFM transition of Ce. For $x \geq 0.75$, the magnetic susceptibility $\chi(T)$ shows a sharp increase at a temperature $T^\textup{Co}_\textup{C}$ that increases with $x$, reaching 75 K at $x = 1$. Moreover, the zero-field cooling (ZFC) and field cooling (FC) data become separated below $T^\textup{Co}_\textup{C}$ [see, for example, the curves of $x$ = 0.75 in Fig.6(a) and $x$ = 1 in Fig.7]. All these indicate that the Co-ions undergo a FM transition at $T^\textup{Co}_\textup{C}$, which value can be obtained from the derivatives of d$\chi$/d$ T $ as shown in the inset of Fig.6(b). In this doping region, the magnetic transition associated with Ce $4f$ electrons becomes indiscernible but one can still determine it after properly subtracting a FM background of Co (see below for details). In the high temperature range, the magnetic susceptibility $\chi(T)$ can be well described by the Curie-Weiss law: $\chi(T) = C/(T - \theta_p)$, with $C$ the Curie constant and $\theta_p$ the paramagnetic Curie temperature. In the inset of Fig.6(a), we plot 1/$\chi(T)$ versus $T$ for $x$ = 0.75 and 1, respectively. Here the red lines show the fits to a Curie-Weiss law. For $x = 0.75$, the effective moment and paramagnetic Curie temperature are derived to be 2.54 $\mu_B$ and -31.2 K, respectively; the former one is very close to the free-ion moment of Ce. On the other hand, the effective moment reaches 2.81 $\mu_B$ for $x = 1$, which is significantly larger than that of the free-ion moment of Ce. According to $\mu_\textup{eff} = {(\mu{^2_\textup{Co}}+ \mu{^2_\textup{Ce}})}^{1/2}$, we can estimate the effective moment of Co, which gives $\mu_\textup{Co}$=1.2 $\mu_B$, being consistent with that reported in Ref.26.

As mentioned above, the magnetic transition of Ce is concealed by the strong FM background of Co for $x \geq 0.85$. In order to track the magnetic transition of Ce in these samples, we need subtract the FM contributions of Co-ions which can be approximated by applying an appropriate polynomial fit to the experimental data in a temperature region below $T^\textup{Co}_\textup{C}$. As an example, we show the magnetic susceptibility $\chi(T)$ (both ZFC and FC) of CeCoAsO in Fig.7. The hump in the ZFC data is likely caused by a FM domain effect. The solid lines represents a polynomial fit to the experimental data between 20 K and 50 K, which can be approximately treated as the FM contributions of Co. The so-subtracted magnetic susceptibility $\Delta \chi$ of Ce-ions is plotted in the inset of Fig.7, which allows us to determine the magnetic transition of Ce from the maximum of $\Delta\chi(T)$ as marked by the arrow in the inset. Here the magnetic transition is again broadened attributed to the possible polarization effect of Co-moments. Nevertheless, the peak structure of $\Delta \chi(T)$ still indicates an AFM-type transition of Ce-moments at $T^\textup{Ce}_\textup{N}$.

Figure 8 presents the field dependence of the magnetization $M(H)$ at various temperatures. For $x\leq 0.7$, $M(H)$ exhibits linear behavior without hysteresis, suggesting the absence of spontaneous magnetization. In CeFeAsO [Fig.8a], a kink was observed around 2 T at $T = $ 2 K which might correspond to a spin-flop transition of Ce.~\cite{luo2010phase} While the Co-ions become ferromagnetically ordered at $x \geq 0.75$, the magnetization $M(H)$ shows an obvious nonsaturating hysteresis loop at temperatures below $T_\textup{C}^\textup{Co}$; its size increases with either increasing Co concentration or decreasing temperature. The large width of the hysteresis loops, and especially the nearly-square loop for $x = 0.9$, indicates that Ce is producing exchange-spring-like behavior in the Co. From Fig.9, one can see that the saturated moments, derived by extrapolating the linear part of the $M(H)$ curves at temperatures near $T_\textup{C}^\textup{Co}$ to zero field, increase with increasing Co
concentration, reaching 0.17$\mu _{B}$ at $x = 1$ (see the inset); The low-temperature remanence in the loops in Fig.8 indicates a saturation moment approximately double that value. It is noted that such a large hysteresis was not observed in other $Re$CoAsO compounds except for CeCoAsO.~\cite{yanagi2008itinerant, mcguire2010magnetic, awana2010magnetic, ohta2009magnetic}

\subsubsection{GdFe$_{1-y}$Co$_y$AsO}

\begin{table}[tbp]
\caption{The superconducting transition temperatures $T_\textup{sc}$ (K) for GdFe$_{1-y}$Co$_y$AsO, $0.07 \leq y \leq 0.2$, determined from 90$\%$ ($T^\protect\rho_\textup{sc}$ (onset)), 50$\%$ ($T^\protect\rho_\textup{sc}$ (mid)) and 10$\%$ ($T^\protect\rho_\textup{sc}$ (zero)) of the normal resistivity at $T_\textup{sc}$ and the onset of magnetic susceptibility $T{^\protect\chi_\textup{sc}}$.\newline}
\label{tab:table2}\centering
\begin{ruledtabular}
\begin{tabular}{lcccc}
\textrm{$y$}&
\textrm{$T{^\rho_\textup{sc}}$ (onset)}&
\textrm{$T{^\rho_\textup{sc}}$ (mid)}&
\textrm{$T{^\rho_\textup{sc}}$ (zero)}&
\textrm{$T{^\chi_\textup{sc}}$}\\
\colrule
0.07 & 14 & 12.6 & 11.3 & 13\\
0.10 & 19 & 17.6 & 17  & 18\\
0.12 & 15 & 13.7 & 12.6 & 15\\
0.15 & 13 & 12.2 & 11 & 11\\
0.20 & 6 & 4.3 & 2.6 &\\
\end{tabular}
\end{ruledtabular}
\end{table}

\begin{figure}[tbp]
\begin{center}
\includegraphics[width=3.35in,keepaspectratio]{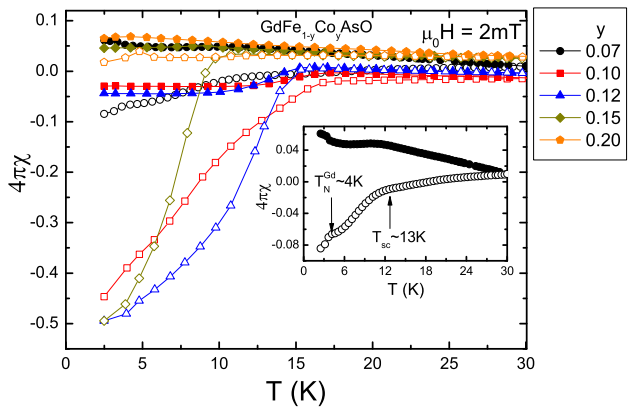}
\end{center}
\caption{(Color online) Low temperature magnetic susceptibility $\protect\chi(T)$ for GdFe$_{1-y}$Co$_y$AsO, $0.07 \leq y \leq 0.20$. The open and filled circles show the ZFC and FC data, respectively. The inset shows the case for $y = 0.07$, where the AFM transition of Gd was observed below the superconducting transition at $T_\textup{sc}$ = 13 K.}
\label{fig10}
\end{figure}

\begin{figure}[tbp]
\begin{center}
\includegraphics[width=3.35in,keepaspectratio]{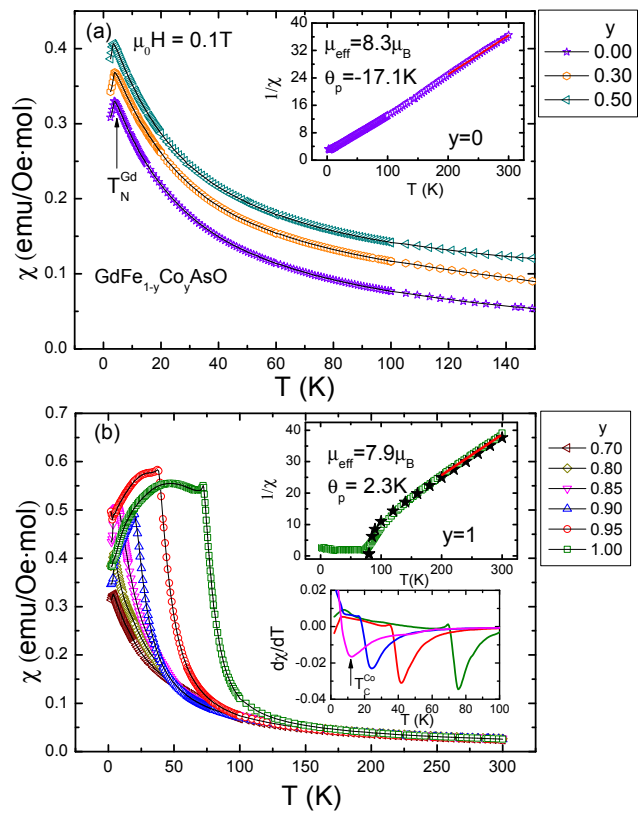}
\end{center}
\caption{(Color online) Temperature dependence of the dc magnetic susceptibility of GdFe$_{1-y}$Co$_y$AsO, $0 \leq y \leq 1$. No difference was observed for the ZFC- and FC-data. Note that the curves for $y = 0.3$ and $y = 0.5$ have been shifted by an offset of 0.05 emu/Oe$\cdot$mol. The insets plot the inverse susceptibility as a function of temperature for $y = 0$ (a) and $y = 1$ (b). The red lines are fits to the Curie-Weiss law and the black stars are a fit to the mean field model. The lower inset of (b) shows d$\protect\chi(T)$/d$T$ versus $T$.}
\label{fig11}
\end{figure}

\begin{figure*}[tbp]
\begin{center}
\includegraphics[width=7in,keepaspectratio]{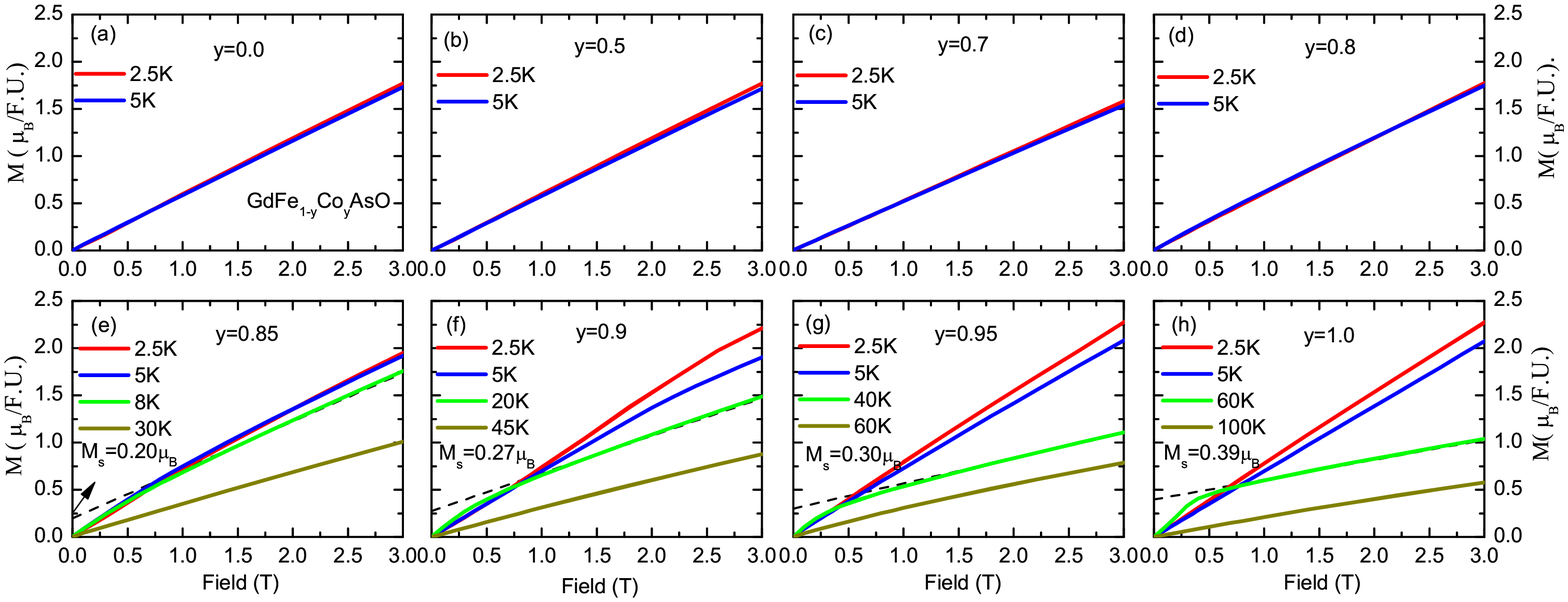}
\end{center}
\caption{(Color online) Field dependence of the magnetization M($H$) at various temperatures for GdFe$_{1-y}$Co$_y$AsO, $0 \leq y \leq 1$.}
\label{fig12}
\end{figure*}

The low temperature magnetic susceptibility $\chi(T)$ for the GdFe$_{1-y}$Co$_y$AsO superconducting samples, is presented in Fig.10. Bulk superconductivity was observed for $0.07 \leq y \leq 0.2$, with the superconducting volume fraction reaching over 50$\%$ for $y$ = 0.1, 0.12 and 0.15. Moreover, a magnetic transition of Gd can be tracked around 4 K as denoted by the arrow in the inset. The superconducting transition temperatures $T_\textup{sc}$, derived from the electrical resistivity and magnetic susceptibility, are summarized in Table II.

Figure 11 plots the magnetic susceptibility of the non-superconducting samples, measured in a field of $\mu _{0}H$ = 0.1 T. To make it clear, a value of 0.05 emu/Oe$\cdot$mol is added as an offset in Fig.11(a). For $x \leq 0.7$, the magnetic susceptibility $\chi(T)$ shows a peak around 4 K, which corresponds to the AFM transition of Gd, and is nearly independent of the Co doping concentration. In the high temperature range ($T >$ 200 K), the magnetic susceptibility $\chi (T)$ can be well described by the Curie-Weiss law. In the inset of Fig.11(a), we plot 1/$\chi (T)$ versus $T$ for GdFeAsO; the red line is a fit to the Curie-Weiss law. The derived effective moment and paramagnetic Curie temperature are 8.3$\mu _{B}$ and -17.1 K, respectively. This effective moment is close to but somewhat lager than the free-ion moment of Gd, possibly due to the contributions of $5d$ electrons.~\cite{moon} As the Fe/Co substitution increases beyond $y=0.7$, the behavior changes dramatically. The susceptibility, as shown in Fig.11(b), rises sharply at a temperature that increases with $y$, reaching 75 K at $y=1$. The inverse susceptibility for $y=1$ can be also fitted by the Curie-Weiss law at temperatures above 200 K [see the upper inset of Fig.11(b)], with a slope consistent with a magnetic moment of 7.9 $\mu _{B}$, very close to the free-ion moment of Gd. However, the value of $\theta _{p}$ = 2.3 K, is substantially small. The downward curvature of the inverse susceptibility is characteristic of a ferrimagnet, in which two magnetic species are coupled antiferromagnetically. The strong dependence of the transition temperature $T_\textup{C}^\textup{Co}$ with Co concentration [see the lower inset of Fig.11(b)] suggests that the FM coupling among Co-atoms, rather than strong AFM coupling to the Gd, is the primary driver of the transition.

We model the behavior within a mean field approach,~\cite{Kittel} in which the magnetization of Co atoms is given by $M_\textup{Co}T = C_\textup{Co} (\mu_0H+\lambda M_\textup{Co}-\mu M_\textup{Gd})$, and that of the Gd by $M_\textup{Gd}T = C_\textup{Gd}(\mu_0H-\mu M_\textup{Co})$. Here $C_\textup{Co}$($C_\textup{Gd}$) is the Curie constant for Co(Gd), $\lambda$ is the FM Co-Co coupling constant and $\mu$ is the AFM Co-Gd coupling. We assume that $C_\textup{Gd}$ follows from the full atomic moment of Gd, 7.9$\mu_B$, and seek a solution in which $T^\textup{Co}_\textup{C} \approx \lambda C_\textup{Co}$ and $\theta_p \approx 2.3$ K. The solid star symbols in the inset of Fig.11(b) show a fit to the data with $\mu / \lambda \approx 0.02$, confirming that the AFM coupling between the two species is relatively weak.

Despite the weak coupling, the large magnetic moment of Gd leads to an appreciable magnetization that opposes that of the Co, causing a strong decrease in the combined magnetization at low temperature. The data in Fig.12 confirm this picture, which plots the field dependence of magnetization at different temperatures with the magnetic field up to 3 T. Below $y = 0.85$, the magnetization is strictly linear in field and independent of Co concentration. For $y\geq 0.85$, in which the Co moments are ferromagnetically ordered, spontaneous magnetization was observed below the Curie temperature $T_\textup{C}^\textup{Co}$ and its size increases with increasing the Co concentration. For GdCoAsO, the magnetization reaches 0.39 $\mu _{B}$ with a maximum near 60 K.  Similar results were also reported in LaCoAsO, NdCoAsO and SmCoAsO compounds with a saturated moment of 0.46$\mu _{B}$, 0.20$\mu _{B}$ and 0.18$\mu _{B}$ respectively.~\cite{yanagi2008itinerant, mcguire2010magnetic, awana2010magnetic} The saturated moment of Co in these compounds is very small, meaning that the ferromagnetism of Co is itinerant in nature. The fit shown in Fig.11, however, suggests that the Co moment is closer to 0.5$\mu_{B}$, indicating that the spontaneous magnetization in Fig.12 is reduced by the negative magnetization induced in the Gd. At low temperatures, the spontaneous magnetization has disappeared and the magnetization is again linear in field. This contrasts sharply with the low temperature hysteretic behavior observed in CeFe$_{1-x}$Co$_{x}$AsO, demonstrating that the admixture of $4f$ and $3d$ electrons dominates in CeCoAsO as compared with the fully localized $4f$electrons of GdCoAsO. We expect that the Gd blocks order antiferromagnetically at low temperature and tend to align the magnetically-ordered Co blocks oppositely. This type of behavior has been observed by means of neutron scattering in NdCoAsO.~\cite{marcinkova2010superconductivity,mcguire2010magnetic} In that case, the AFM interactions among the Nd atoms and between Nd and Co are sufficiently strong to reorient the FM Co layers into an antiparallel alignment.

\subsection{Heat Capacity}

\subsubsection{CeFe$_{1-x}$Co$_x$AsO}

\begin{figure}[tbp]
\begin{center}
\includegraphics[width=3.2in,keepaspectratio]{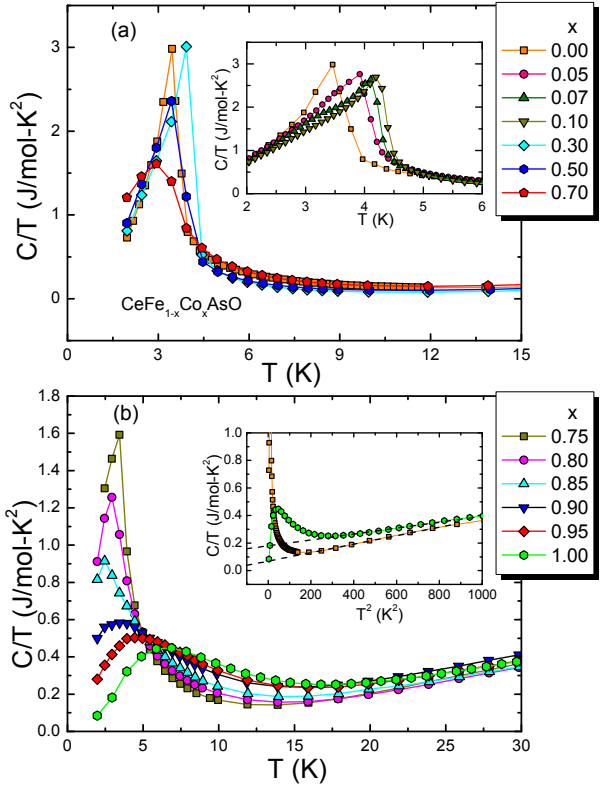}
\end{center}
\caption{(Color online) Temperature dependence of the specific heat $C(T)/T$ at zero field for CeFe$_{1-x}$Co$_x$AsO, $0 \leq x \leq 1$. Inset of (a)
shows the data for low Co concentrations ($0 \leq x \leq 0.1$). Inset of (b) plots $C(T)/T$ versus $T^2$ for CeFeAsO and CeCoAsO. The dashed lines are fits to $C/T = \protect\gamma_0 + \protect\beta T^2$.}
\label{fig13}
\end{figure}

In order to further characterize the magnetic order of Ce $4f$ electrons in CeFe$_{1-x}$Co$_{x}$AsO, we also measured the temperature dependence of
specific heat $C(T)$. Figure 13 presents the low-temperature specific heat $C(T)/T$ of several representative doping concentrations. One can see that the specific heat $C(T)/T$ shows a $\lambda $-type transition for $x\leq 0.8$, characteristic of a typical AFM transition. The transition temperatures are consistent with those derived from the electrical resistivity and magnetic susceptibility as shown above. In previous measurements of neutron scattering and muon spin relaxation, it was found that there exists a strong coupling between Ce-$4f$ and Fe $3d$ electrons in CeFeAsO.~\cite{chi2008crystalline, maeter2009interplay} In this case, suppression of the SDW transition of Fe may have a significant impact on the magnetic order of Ce. To confirm it, we have performed a detailed measurement of low temperature specific heat for $x\leq 0.1$. As seen in the inset of Fig.13(a), the AFM transition of Ce indeed shifts to higher temperature while suppressing the SDW transition of Fe by Fe/Co substitution, indicating a kind of competing interaction between Ce $4f$ and Fe $3d$ electrons. It is noted that, in the electrical resistivity $\rho (T)$, the SDW transition is barely visible for $x>0.06$. However, the continuous increase of $T_\textup{N}^\textup{Ce}$ up to $x=0.1$ may indicate that the critical doping concentration for suppressing the SDW transition is around $x\simeq 0.1$, which is compatible if we extrapolate the SDW-phase boundary in the phase diagram as shown in Fig.15(a). Furthermore, the specific heat $C(T)/T$ of the superconducting samples shows a pronounced magnetic transition of Ce around $T_\textup{N}^\textup{Ce}\approx $ 4 K. One notes that the specific heat jump $\Delta C/T$ at the superconducting transition (about 90 mJ/mol-K$^2$ assuming $\Delta C/T_\textup{sc}=1.43\gamma_0$) is too tiny to be observed in comparison with that at $T_\textup{N}^\textup{Ce}$.

With further increase of Co content, the AFM transition of Ce depends weakly on the doping concentration up to $x=0.8$. Above that, the transition is significantly broadened and its maximum is slightly shifted towards higher temperatures with increasing $x$, being similar to the resistive results. Such behavior is likely attributed to the Co-induced polarization on Ce-moments, which may take place prior to the AFM order of Ce at $T_\textup{N}^\textup{Ce}$. The specific heat data can be fitted by $C/T=\gamma _{0}+\beta T^{2}$ at temperatures above the magnetic transition of Ce [see the dashed lines in the inset of Fig.13(b)], from which one can estimate the Sommerfeld coefficient $\gamma _{0}$. The derived $\gamma _{0}$ value is summarized in the inset of Fig.15, which is nearly a constant for $0\leq x\leq 0.7$, but then increases with further increasing $x$. In comparison with CeFeAsO ($\gamma _{0}$ $\simeq$ 65 mJ/mol-K$^2$), the $\gamma _{0}$ value of CeCoAsO is significantly enhanced ($\gamma _{0} \simeq $ 186 mJ/mol-K$^{2}$), implying an increase of the hybridization between $4f$ and $3d$ electrons upon substituting Fe with Co. The large electronic specific heat coefficient and the strong temperature dependence of the electrical resistivity [see Fig.2(c)] suggest strong electron correlations on the Co-rich side. We note that the $\gamma _{0}$ value of CeCoAsO is also much larger than that of LaCoAsO ($\gamma _{0} \approx$ 20 mJ/mol-K$^2$),~\cite{anand2011} excluding the possibility that the enhanced $\gamma _{0}$ value is due to the magnetic scattering of Co.

\subsubsection{GdFe$_{1-y}$Co$_y$AsO}

\begin{figure}[tbp]
\begin{center}
\includegraphics[width=3.35in,keepaspectratio]{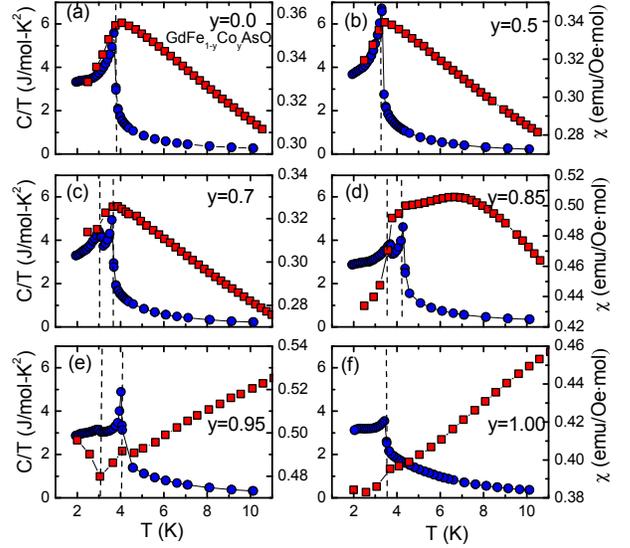}
\end{center}
\caption{(Color online) Temperature dependence of the specific heat $C(T)$ (blue circle, left axis) and magnetic susceptibility $\protect\chi(T)$ (red square, right axis) for GdFe$_{1-y}$Co$_y$AsO, $0 \leq y \leq 1$.}
\label{fig14}
\end{figure}

We also performed measurements of the low temperature specific heat for GdFe$_{1-y}$Co$_y$AsO. Figure 14 shows the magnetic susceptibility $\chi(T)$ and the specific heat $C(T)/T$ of several representative doping concentrations. For $y < 0.7$, observations of a $\lambda$-type anomaly in specific heat and a cusp in magnetic susceptibility unambiguously identify the AFM transition of Gd around $T_\textup{N}^\textup{Gd} \approx 4$ K. With increasing the Co content, a second anomaly develops in the heat capacity at lower temperature for $y \geq 0.7$, but disappears again at $y = 1$. Similar evidence can be also inferred from the magnetic susceptibility $\chi(T)$, which is most clearly seen in $y = 0.85$. As already described above, the second transition was also observed in the electrical resistivity [see Fig.3(b)]. All these indicate that the system ($0.7 \leq y < 1.0$) may undergo a magnetic reorientation transition below $T_\textup{N}^\textup{Gd}$ attributed to the coupling between the Co $3d$ and Gd $4f$ electrons.

\subsection{Phase Diagram and Discussion}

\begin{figure}[tbp]
\begin{center}
\includegraphics[width=3.2in,keepaspectratio]{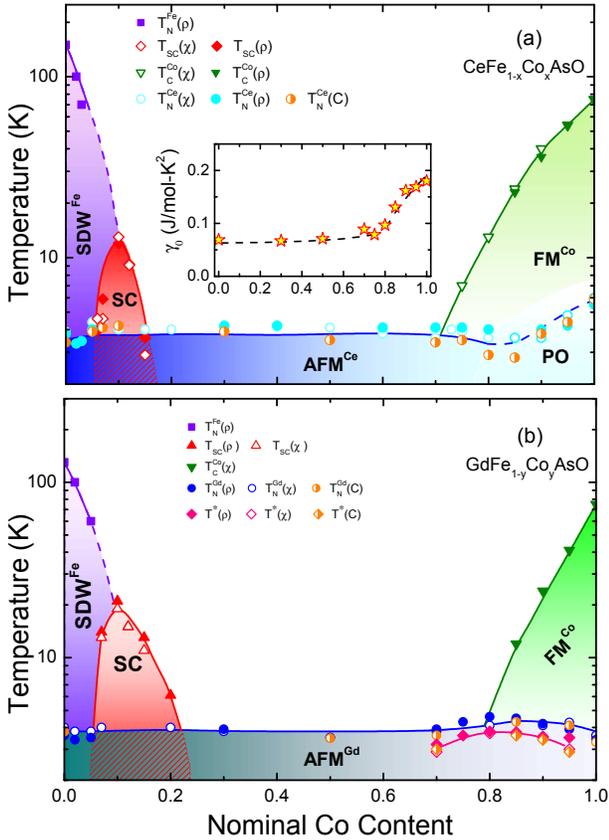}
\end{center}
\caption{(Color online) The magnetic and superconducting phase diagram of CeFe$_{1-x}$Co$_x$AsO (a) and GdFe$_{1-y}$Co$_y$AsO (b), plotted as a function of nominal Co content. Various symbols denote different types of transition temperatures. The lines are guides to the eyes. The inset of (a) plots the Sommerfeld coefficient $\protect\gamma_0$ as a function of Co concentration for CeFe$_{1-x}$Co$_x$AsO.}
\label{fig15}
\end{figure}

Based on the above experimental data, in Fig.15 we present a magnetic and superconducting phase diagram for CeFe$_{1-x}$Co$_{x}$AsO and GdFe$_{1-y}$Co$_{y}$AsO ($0\leq x,y\leq 1$). CeFeAsO is a bad metal, in which Fe $3d$ and Ce $4f$ electrons are antiferromagnetically ordered below 150 K and 4 K, respectively. Upon substituting Fe with Co, the magnetic order of Fe $3d$ electrons is suppressed around $x \approx 0.1$ (marked by SDW$^\textup{Fe}$) and superconductivity appears in a narrow range of $0.05 < x < 0.20$ (marked by SC), indicating that the formation of superconductivity is closely related to the magnetism of Fe in these compounds. At lower temperatures, Ce $4f$ electrons undergo an AFM transition in the entire doping-range (marked by AFM$^\textup{Ce}$); the N\'{e}el temperature $T_\textup{N}^\textup{Ce}$ slightly increases with suppressing the SDW transition of Fe and then remains nearly unchanged up to $x = 0.8$. With further increasing Co-content ($x\geq 0.75$), the Co $3d$ electrons form long range FM order, the Curie temperature of which eventually increases with $x$, reaching $T_\textup{C}^\textup{Co}\simeq 75$ K in CeCoAsO (marked by FM$^\textup{Co}$). On the Co-rich side, the Co-ferromagnetism has a strong polarization effect on the AFM order of Ce (marked by PO). GdFeAsO is also a bad metal, with the Fe $3d$ and Gd $4f$ electrons ordered antiferromagnetically below 130 K and 4 K, respectively. The Fe-SDW transition vanishes around $y = 0.1$ and superconductivity appears in a range of $0.05 < y < 0.25$. At lower temperatures, Gd is antiferromagnetically ordered with $T_\textup{N}^\textup{Gd}$ $\approx $ 4 K (marked by AFM$^\textup{Gd}$), being nearly independent of the Fe/Co substitution. Meanwhile, long range FM order of Co $3d$ electrons develops above $y\approx 0.8$, with the transition temperature reaching 75 K in GdCoAsO. Unlike CeCoAsO, however, GdCoAsO shows ferrimagnetic behavior, with clear evidence that the Gd moments oppose that of Co, leading to a strong suppression of the magnetization at low temperatures. Moreover, for $0.7 \leq y <1.0$, it appears that a magnetic reorientation transition occurs below $T_\textup{N}^\textup{Gd}$, as noted by $T^\ast$ in Fig.15(b).

As shown above, the 1111-type iron pnictides exhibit remarkably rich magnetic and superconducting properties via elemental substitutions. Different from CeFe$_{1-x}$Co$_{x}$AsO and GdFe$_{1-y}$Co$_{y}$AsO, superconductivity was only observed around $x = 0.3$ in CeFeAs$_{1-x}$P$_{x}$O, which seems to coexist with ferromagnetism of Ce below 4 K.~\cite{Jesche} The magnetic order of Ce persists up to $x = 0.9$ in CeFeAs$_{1-x}$P$_{x}$O, but it changes from an AFM- to a FM-type order while the SDW transition of Fe is suppressed around $x \simeq $ 0.3.~\cite{luo2010phase, de2010lattice, Jesche} In contrast, the AFM order of Ce-ions is remarkably robust against Fe/Co substitution. Its ordering temperature $T_\textup{N}^\textup{Ce}$ increases with $x$ while suppressing the SDW transition of Fe, followed by a nearly unchanged value. The persistence of the Ce-AFM order, with a slight dependence of $T_\textup{N}^\textup{Ce}$ on Co-doping, seems to contradict the Kondo physics, which effect would be enhanced when the interlayer distance is compressed by the Fe/Co substitution as evidenced from the enhancement of the Sommerfeld coefficient in CeCoAsO ($\gamma _{0}$ $\sim $ 200 mJ/mol-K$^{2}$).~\cite{sarkar2010interplay} This is possibly due to the polarization of Co-atoms caused by spin-orbit coupling which competes with the Kondo effect. The resulting AFM exchange interaction between Co $3d$ and Ce $4f$ electrons is thus anisotropic in the spin space. The multi-orbital characteristics may also add complications to the interplay among various intralayer and interlayer exchange interactions, leading to diverse behavior in the $3d$-$4f $ electron pnictides. For example, in other $Re$Co$Pn$O systems, e.g., Re = Nd and Sm,\cite{mcguire2010magnetic,ohta2009magnetic,anand} the interlayer $d$-$f$ exchange interaction is strong enough to realign the Co moments, leading to a subsequent AFM transition of Co at a lower temperature. Such transitions and magnetic structures have recently been proposed from neutron scattering in NdCoAsO.\cite{marcinkova2010superconductivity, mcguire2010magnetic} However, in CeFe$_{1-x}$Co$_{x}$AsO, Ce-ions carry a small magnetic moment and the material shows a relatively weak lattice shrinkage along the $c$-axis upon substituting Fe with Co; the resulting AFM exchange interaction is likely not sufficiently strong to realign the Co moments. Instead, the Ce moments are partially polarized by the internal field of Co. However, the large moment of Gd in GdFe$_{1-y}$Co$_{y}$AsO is robust against the FM order of Co. The coupling between the Gd and Co lattices leads to ferrimagnetic behavior and also possible magnetic reorientation of Co at low temperatures. Since the FM order of Co is generally observed in the $Re$CoAsO families,~\cite{ohta2009magnetic} it is desirable to extend such measurements to other series, to study the interplay of $4f$  and $3d$ electrons and its influence on superconductivity and magnetism in iron pnictides. Further investigations on the Co-rich side is highly important in order to clarify the magnetic structures of $3d$ and $4f$ electrons in these systems.

\section{CONCLUSION}

In summary, we have systematically studied the transport, magnetic and thermodynamic properties of the polycrystalline samples CeFe$_{1-x}$Co$_{x}$AsO and GdFe$_{1-y}$Co$_{y}$AsO ($0 \leq x,y \leq 1$) and presented a complete doping-temperature phase diagram. It is found that both Ce- and Gd-compounds exhibit similar behavior on the Fe-side, but rather distinct properties on the Co-side. On the Fe-rich side, superconductivity occurs upon suppressing the SDW transition of Fe. At lower temperatures, the rare-earth elements (Ce and Gd) undergo an AFM-type magnetic transition, whose temperature is slightly increased while suppressing the SDW-transition of Fe, but then becomes nearly unchanged over a wide doping regime. On the Co-rich side, Co electrons form a long-range FM order, the transition temperature of which increases with increasing Co content and reaches $T_\textup{C}^\textup{Co}\approx $ 75 K in CeCoAsO and GdCoAsO. Evidence of a strong polarization effect of the Co-ferromagnetism on Ce atoms is observed on the Co-rich side. But the Gd is rather robust against Fe/Co substitution; it is hardly polarized even in the case that Co-ions are ferromagnetically ordered. However, the coupling between Gd- and Co-species gives rise to ferrimagnetic behavior on the Co-side, which can be well described in terms of a mean field model that considers weak AFM coupling between the Gd-species and the Co-species. Our findings suggest that, the delicate interplay of $3d$-$4f$ electrons rising from the competition of electron hybridizations and magnetic exchange coupling leads to remarkably rich physics in the iron pnictides.

\begin{acknowledgments}
We would like to thank C. Geibel, E. D. Bauer, R. E. Baumbach, F. C. Zhang, Z. A. Xu, and G. H. Cao for useful discussions. Work at Zhejiang University is supported by the National Basic Research Program of China (Nos.2009CB929104 and 2011CBA00103),the National Science Foundation of China (Nos.10934005, 11174245, 11274084), Zhejiang Provincial Natural Science Foundation of China and the Fundamental Research Funds for the Central Universities. Work at Los Alamos National Lab was performed under the auspices of the US DOE and supported in part by the Los Alamos LDRD program. Work at The University of Texas at Dallas is supported by the AFOSR grant (No.FA9550-09-1-0384).
\end{acknowledgments}


\begin{thebibliography}{10}
\bibitem{kamihara2008iron} Y. Kamihara, T. Watanabe, M. Hirano, and H. Hosono, J. Am. Chem. Soc. \textbf{130}, 3296
    (2008).

\bibitem{cxh2008} X. H. Chen, T. Wu, G. Wu, R. H. Liu, H. Chen, and D. F. Fang, Nature \textbf{453}, 761 (2008).

\bibitem{chen2008superconductivity} G. F. Chen, Z. Li, D. Wu, G. Li, W. Z. Hu, J. Dong, P. Zheng, J. L. Luo, and N. L. Wang, Phys. Rev. Lett. \textbf{100}, 247002 (2008).

\bibitem{Ren2008} Z. A. Ren, J. Yang, W. Lu, W. Yi, X. L. Shen, Z. C. Li, G. C. Che, X. L. Dong, L. L. Sun, F. Zhou and Z. X. Zhao, Europhys. Lett. \textbf{82}, 57002 (2008).

\bibitem{wang2008thorium} C. Wang, L. J. Li, S. Chi, Z. W. Zhu, Z. Ren, Y. Y. Li, Y. T. Wang, X. Lin, Y. K. Luo, S. Jiang, X. F. Xu, G. H. Cao, and Z. A. Xu, Europhys. Lett. \textbf{83}, 67006 (2008).

\bibitem{armitage2010progress} N. P. Armitage, P. Fournier, and R. L. Greene, Rev. Mod. Phys. \textbf{82}, 2421 (2010).

\bibitem{reviews} See reviews, e.g., H. H. Wen, Adv. Mater. \textbf{20}, 3764 (2008); M. V. Sadovskii, Physics-Uspekhi \textbf{51}, 1201 (2009); J. Paglione and R. L. Greene, Nature physics \textbf{6}, 645 (2010).

\bibitem{mazin2008unconventional} I. I. Mazin, D. J. Singh, M. D. Johannes and M. H. Du, Phys. Rev. Lett. \textbf{101}, 057003 (2008).

\bibitem{Mazinreview} I. I. Mazin and J. Schmalian, Physica C \textbf{469}, 614 (2009).

\bibitem{Harlingen} D. J. Van Harlingen, Rev. Mod. Phys. \textbf{67}, 515 (1995).

\bibitem{Tsuei} C. C. Tsuei and J. R. Kirtley, Rev. Mod. Phys. \textbf{72}, 969 (2000).

\bibitem{yuan2009nearly} H. Q. Yuan, J. Singleton, F. F. Balakirev, S. A. Baily, G. F. Chen, J. L. Luo and  N. L. Wang, Nature \textbf{457}, 7229 (2009).

\bibitem{ZhangHC2} J. L. Zhang, L. Jiao, Y. Chen and H. Q. Yuan, Front. Phys. \textbf{6}, 463, (2011).

\bibitem{worthington1987anisotropic} T. K. Worthington, W. J. Gallagher and T. R. Dinger, Phys. Rev. Lett. \textbf{59}, 1160 (1987).

\bibitem{zhao2010effects} L. D. Zhao, D. Berardan, C. Byl, L. P. Gaudart, and N. Dragoe, J. Phys.: Condens. Matter \textbf{22}, 115701 (2010).

\bibitem{sefat2008superconductivity} A. S. Sefat, A. Huq, M. A. McGuire, R. Jin, B. C. Sales, D. Mandrus,
L. M. D. Cranswick, P. W. Stephens, and K. H. Stone, Phys. Rev. B \textbf{78}, 104505 (2008).

\bibitem{shirage2009search} P. M. Shirage, K. Miyazawa, H. Kito, E. H. Isaki, and A. Iyo, Physica C: Superconductivity  \textbf{469}, 898 (2009).

\bibitem{marcinkova2010superconductivity}  A. Marcinkova, D. A. M. Grist, I. Margiolaki, T. C. Hansen, S. Margadonna,
and J. W. G. Bos, Phys. Rev. B \textbf{81}, 064511 (2010).

\bibitem{wang2009effects} C. Wang, Y. K. Li, Z. W. Zhu, S. Jiang, X. Lin, Y. K. Luo, S. Chi, L. J. Li, Z. Ren,
M. He, H. Chen, Y. T. Wang, Q. Tao, G. H. Cao, and Z. A. Xu, Phys. Rev. B \textbf{79}, 054521 (2009).

\bibitem{yang2008superconductivity} J. Yang, Z. C. Li, W. Lu, W. Yi, X. L. Shen, Z. A. Ren, G. C. Che, X. L. Dong, L. L. Sun, F. Zhou, and Z. X. Zhao,  Supercond. Sci. Technol. \textbf{21}, 082001 (2008).

\bibitem{zhao2008structural} J. Zhao, Q. Huang, C. de La Cruz, S. Li, J. W. Lynn, Y. Chen, M. A. Green, G. F. Chen, G. Li, Z. Li, J. L. Luo, N. L. Wang, and P. C. Dai, Nature Materials \textbf{7}, 953 (2008).

\bibitem{E.M.Bruning2008cefepo} E. M. Br\"{u}ning, C. Krellner, M. Baenitz, A. Jesche, F. Steglich,
and C. Geibel, Phys. Rev. Lett. \textbf{101}, 117206 (2008).

\bibitem{luo2010phase} Y. K. Luo, Y. K. Li, S. Jiang, J. H. Dai, G. H. Cao, and Z. A. Xu, Phy. Rev. B \textbf{81}, 134422 (2010).

\bibitem{de2010lattice} C. de La Cruz, W. Z. Hu, S. Li, Q. Huang, J. W. Lynn, M. A. Green, G. F. Chen, N. L. Wang, H. A. Mook, Q. M. Si, and P. C. Dai, Phys. Rev. Lett. \textbf{104}, 017204 (2010).

\bibitem{Jesche} A. Jesche, T. F\"{o}rster, J. Spehling, M. Nicklas, M. de Souza, R. Gumeniuk, H. Luetkens, T. Goltz, C. Krellner, M. Lang, J. Sichelschmidt, H. H. Klauss, and C. Geibel, Phys. Rev. B \textbf{86}, 020501 (2012).


\bibitem{sarkar2010interplay} R. Sarkar, A. Jesche, C. Krellner, M. Baenitz, C. Geibel, C. Mazumdar, and A. Poddar, Phys. Rev. B \textbf{82}, 054423 (2010).

\bibitem{krellner2009interplay} C. Krellner, U. Burkhardt, and C. Geibel, Physica B: Condensed Matter  \textbf{404}, 3206 (2009).

\bibitem{yanagi2008itinerant} H. Yanagi, R. Kawamura, T. Kamiya, Y. Kamihara,  M. Hirano, T. Nakamura, H. Osawa, and H. Hosono, Phys. Rev. B \textbf{77}, 224431 (2008).

\bibitem{awana2010magnetic} V. P. S. Awana, I. Nowik, A. Pal, K. Yamaura, E. T. Muromachi, and I. Felner, Phys. Rev. B \textbf{81}, 212501 (2010).

\bibitem{mcguire2010magnetic} M. A. McGuire, D. J. Gout, V. O. Garlea, A. S. Sefat, B. C. Sales, and D. Mandrus, Phys. Rev. B \textbf{81}, 104405 (2010).

\bibitem{ohta2009magnetic} H. Ohta, and K. Yoshimura, Phys. Rev. B \textbf{80}, 184409 (2009).

\bibitem{dong2008competing} J. Dong, H. J. Zhang, G. Xu, Z. Li, G. Li, W. Z. Hu, D. Wu, G. F. Chen, X. Dai, J. L. Luo, Z. Fang and N. L. Wang, Europhys. Lett. \textbf{83}, 27006 (2008).

\bibitem{tianCe} T. Shang, L. Jiao, J. H. Dai, H. Q. Yuan, F. F. Balakirev, W. Z. Hu and L. N. Wang, arxiv: 1209.4529.

\bibitem{CeRh2Si2} S. Araki, M. Nakashima, R. Settai, T. C. Kobayashi and Y. Onuki, J. Phys.: Condens. Matter \textbf{14}, L377 (2002).

\bibitem{Ce2Rh3Ge5} Z. Hossain, H. Ohmoto, K. Umeo, F. Iga, T. Suzuki, T. Takabatake, N. Takamoto and K. Kindo,  Phys. Rev. B \textbf{60}, 10383 (1999).

\bibitem{jiao} L. Jiao, Y. Kohama, J. L. Zhang, H. D. Wang, B. Maiorov, F. F. Balakirev, Y. Chen, L. N. Wang, T. Shang, M. H. Fang and H. Q. Yuan, Phys. Rev. B \textbf{85}, 064513 (2012).

\bibitem{aroszynski} J. Jaroszynski, F. Hunte, L. Balicas, Y. J. Jo, I. Rai\v{c}evi\'{c}, A. Gurevich, D. C. Larbalestier, F. F. Balakirev
L. Fang, P. Cheng, Y. Jia, and H. H. Wen, Phys. Rev. B \textbf{78}, 174523 (2008).

\bibitem{Lee} H. S. Lee, M. Bartkowiak,  J. S. Kim and H. J. Lee, Phys. Rev. B \textbf{82}, 104523 (2010).

\bibitem{werthamer1966temperature} N. R. Werthamer, E. Helfand, and P. C. Hohenberg, Phys. Rev. \textbf{147}, 295 (1966).

\bibitem{moon} R. M. Moon, W. C. Koehler, J. W. Cable, and H. R. Child, Phys. Rev. B \textbf{5}, 997 (1972).

\bibitem{Kittel} C. Kittel, \emph{Introduction to Solid State Physics}, (Wiley, New York, 1971)

\bibitem{chi2008crystalline} S. Chi, D. T. Adroja, T. Guidi, R. Bewley, S. Li, J. Zhao, J. W. Lynn, C. M. Brown, Y. Qiu, G. F. Chen, J. L. Lou, N. L. Wang, and P. C. Dai, Phys. Rev. Lett. \textbf{101}, 217002 (2008).

\bibitem{maeter2009interplay} H. Maeter, H. Luetkens, Y. G. Pashkevich, A. Kwadrin, R. Khasanov, A. Amato, A. A. Gusev, K. V. Lamonova, D. A. Chervinskii, R. Klingeler, C. Hess, G. Behr, B. B\"{u}chner, and H. H. Klauss, Phys. Rev. B \textbf{80}, 094524 (2009).

\bibitem{anand2011}A. Pal, M. Tropeano, S. D. Kaushik, M. Hussain, H. Kishan, and V. P. S. Awana, J. Appl. Phys. \textbf{109}, 071121 (2011)

\bibitem{anand} A. Pal, S. S. Mehdi, M. Hussain, B. Gahtori and V. P. S. Awana, J. Appl. Phys. \textbf{110}, 103913 (2011).

\end{thebibliography}
\end{document}